%% file: paper.tex
\documentclass[aps,prl,twocolumn,superscriptaddress]{revtex4-1}
\pdfoutput=1
\usepackage{graphicx}
\usepackage{color}
\usepackage{hyperref}

\newcommand{\apjl}{ApJL}
\newcommand{\araa}{ARAA}

\newcommand{\mnras}{MNRAS}
\newcommand{\pasp}{PASP}
\newcommand{\physrep}{Phys.~Rep.}
\newcommand{\redm}{redMaPPer}
\newcommand{\redms}{redMaPPer }
\newcommand{\mtom}{M_{\rm 200m}}

\newcommand{\ctom}{c_{\rm 200m}}
\def\ave#1{\left\langle #1 \right\rangle}

\input{commands}

\begin{document}


\title{Evidence of Halo Assembly Bias in Massive Clusters}


\author{Hironao Miyatake}
\email{Hironao.Miyatake@jpl.nasa.gov}
\affiliation{Department of Astrophysical Sciences, Princeton University,
Peyton Hall, Princeton NJ 08544, USA}
\affiliation{Kavli Institute for the Physics and Mathematics of the
Universe (WPI),
UTIAS,
The University of
Tokyo, Chiba, 277-8583, Japan}
\affiliation{Jet Propulsion Laboratory, California Institute of Technology, Pasadena, CA 91109, USA}
\author{Surhud More}
\affiliation{Kavli Institute for the Physics and Mathematics of the
Universe (WPI),
UTIAS,
The University of
Tokyo, Chiba, 277-8583, Japan}
\author{Masahiro Takada}
\affiliation{Kavli Institute for the Physics and Mathematics of the
Universe (WPI),
UTIAS,
The University of
Tokyo, Chiba, 277-8583, Japan}
\author{David N. Spergel}
\affiliation{Department of Astrophysical Sciences, Princeton University,
Peyton Hall, Princeton NJ 08544, USA}
\affiliation{Kavli Institute for the Physics and Mathematics of the
Universe (WPI),
UTIAS,
The University of
Tokyo, Chiba, 277-8583, Japan}
\author{Rachel Mandelbaum}
\affiliation{McWilliams Center for Cosmology, Department of Physics,
Carnegie Mellon University, Pittsburgh, PA 15213, USA}
\author{Eli S. Rykoff}
\affiliation{Kavli Institute for Particle Astrophysics \& Cosmology, P. O. Box 2450, Stanford University, Stanford, CA 94305, USA}
\affiliation{SLAC National Accelerator Laboratory, Menlo Park, CA 94025, USA}
\author{Eduardo Rozo}
\affiliation{Department of Physics, University of Arizona, 1118 E 4th
St, Tucson, AZ 85721, USA}


\date{\today}

\begin{abstract}
 We present significant evidence of halo assembly bias for SDSS
 \redms galaxy clusters in the redshift range $[0.1, 0.33]$.  By
 dividing the 8,648 clusters into two subsamples based on the average
 member galaxy separation from the cluster center, we first show that
 the two subsamples have very similar halo mass of $M_{200m}\simeq
 1.9\times 10^{14}~h^{-1}M_\odot$ based on the weak lensing signals at
 small radii $R\simlt 10~h^{-1}{\rm Mpc}$. However, their halo bias
 inferred from both the large-scale weak lensing and the projected
 auto-correlation functions differs by a factor of $\sim$1.5, which is a
 signature of assembly bias.  The same bias hypothesis for the two
 subsamples is excluded at 2.5$\sigma$ in the weak lensing and
 4.4$\sigma$ in the auto-correlation data, respectively. This
 result could bring a significant impact on both galaxy evolution and
 precision cosmology.
 \end{abstract}


\maketitle

Since massive cluster-sized halos of cold dark matter (CDM) emerge from
the rarest peaks in the primordial Gaussian random density field
\citep{PressSchechter:1974,Kravtsov:2012}, their clustering amplitudes at
large scales is highly biased compared to the underlying mass
distribution \citep{Kaiser:1984,Mo:1996,Sheth:2001,Tinker:2010}. In the
standard version of the halo model, the clustering amplitude depends
{\it only} on halo mass \citep[see e.g.,][]{Cooray:2002}.

Do secondary parameters other than the halo mass affect the clustering
amplitude? Do observations of galaxy clusters reveal halo assembly bias, an
effect predicted both by analytical theory and simulations \citep{Sheth:2004,
Gao:2005,*Gao:2007, Wechsler:2006, Li:2008, Dalal:2008}?

In this {\it Letter}, we use a combination of the clustering and the
weak gravitational lensing of clusters and present the {\it first}
significant evidence of a difference in the large scale bias for
cluster samples of the same mass.  We divide our sample of galaxy
clusters into two based on the average projected radial separation of
member galaxies, and investigate the difference in their clustering
amplitude on large scales.  In Ref.~\cite{More:2015}, we will show
that the two cluster subsamples have different mass accretion rates
and hence different assembly histories, confirming these observations
to be strong evidence for halo assembly bias. The distinguishing
feature of our analysis is to use weak lensing to verify that the
subsamples have similar halo masses, but different halo biases. There
have been several claims for the evidence of assembly bias on galaxy
scales \cite[e.g.,][]{Yang:2006, Tinker:2012, Hearin:2014}. However,
Ref.~\cite{Linetal:2015} found that the difference in clustering
properties could all be explained as due to difference in halo mass or
contamination by satellite galaxies and concluded that there was no
significant evidence of the assembly bias for galaxy-scale halos.

Throughout this {\it Letter}, we adopt a flat $\Lambda$CDM
cosmological model with matter density parameter $\Omega_{\rm m}=0.27$
and the Hubble parameter $h=0.7$.

{\it Cluster subsamples} -- We use the publicly available catalog of
galaxy clusters identified from the SDSS DR8 photometric galaxy catalog by the {\it
red}-sequence {\it Ma}tched-filter {\it P}robabilistic {\it Per}colation
(\redm) cluster finding algorithm \citep[v5.10 at 
\footnote{\url{http://risa.stanford.edu/redmapper/}}, also see Refs.][for details]{Rykoff:2014,Rozoetal:2014}.  \redms uses the $ugriz$ magnitudes
and their errors, to group spatial concentrations of red-sequence
galaxies at similar redshifts into clusters. For each cluster, the
catalog contains an optical richness estimate $\lambda$, a photometric
redshift estimate $z_\lambda$, as well as the position and probabilities
of 5 candidate central galaxies $p_{\rm cen}$ \citep{Rykoff:2014}.  A separate member galaxy
catalog provides a list of members for each cluster, each of which is
assigned a membership probability, $p_{\rm mem}$ \citep{Rykoff:2014}.

We use a sample of clusters with $20<\lambda<100$ and $0.1\le z_{\lambda}\le
0.33$.  The richness cuts ensure a pure and statistically meaningful sample of
clusters at all richness bins, while the redshift cuts select a nearly
volume-limited sample of clusters \cite{Rykoff:2014}, resulting in a sample of
8,648 clusters. For the weak lensing and clustering measurements, we
use 100 times as many random points as real clusters, incorporating the
survey geometry, depth variations, and distributions of cluster redshift and
richness \citep[see][for details on the use of random catalogs]{Mandelbaum:2005,Miyatake:2015}.

As a proxy for the assembly history of the clusters, we use the average
projected separation of member galaxies from the cluster center,
$\ave{R_{\rm mem}}$. For each cluster, we compute
\begin{equation}
        \ave{R_{{\rm mem}}}={\sum_{i}p_{{\rm mem},i} R_{{\rm
        mem},i}}{\Big/}{\sum_i
                p_{{\rm mem},i}},
  \label{eq:aveR}
\end{equation}
where the summation runs over all member galaxies, and $R_{i}$ is the
projected separation of the $i$-th member from the cluster
center. Throughout this {\it Letter} we use the position of the most
probable central galaxy in each cluster region as a proxy of the
cluster center. We employ 14 equally spaced bins both in redshift and
$\lambda$ and obtain a spline fit for the median of $\ave{R_{\rm
    mem}}$ as a function of redshift and richness.  We then define the
two subsamples by the upper and lower halves of clusters in each bin
of richness and redshift space \footnote{See Supplementary figure for
  the definitions of the two subsamples in the redshift and richness
  plane}. The ratio of $\ave{R_{\rm mem}}$ for a bootstrapped
  realization of galaxy cluster pairs from the large- and the
  small-$\ave{R_{\rm mem}}$ samples selected within the same richness
  and redshift has a distribution with median $1.18^{+0.14}_{-0.09}$
  (the errorbars correspond to the 16th and 84th percentile).  The
large- and small-$\ave{R_{\rm mem}}$ subsamples consist of 4,235 and
4,413 clusters, respectively. By construction, the two subsamples have
almost identical distributions of redshift and richness.

\begin{figure}
 \includegraphics[width=0.5\textwidth]{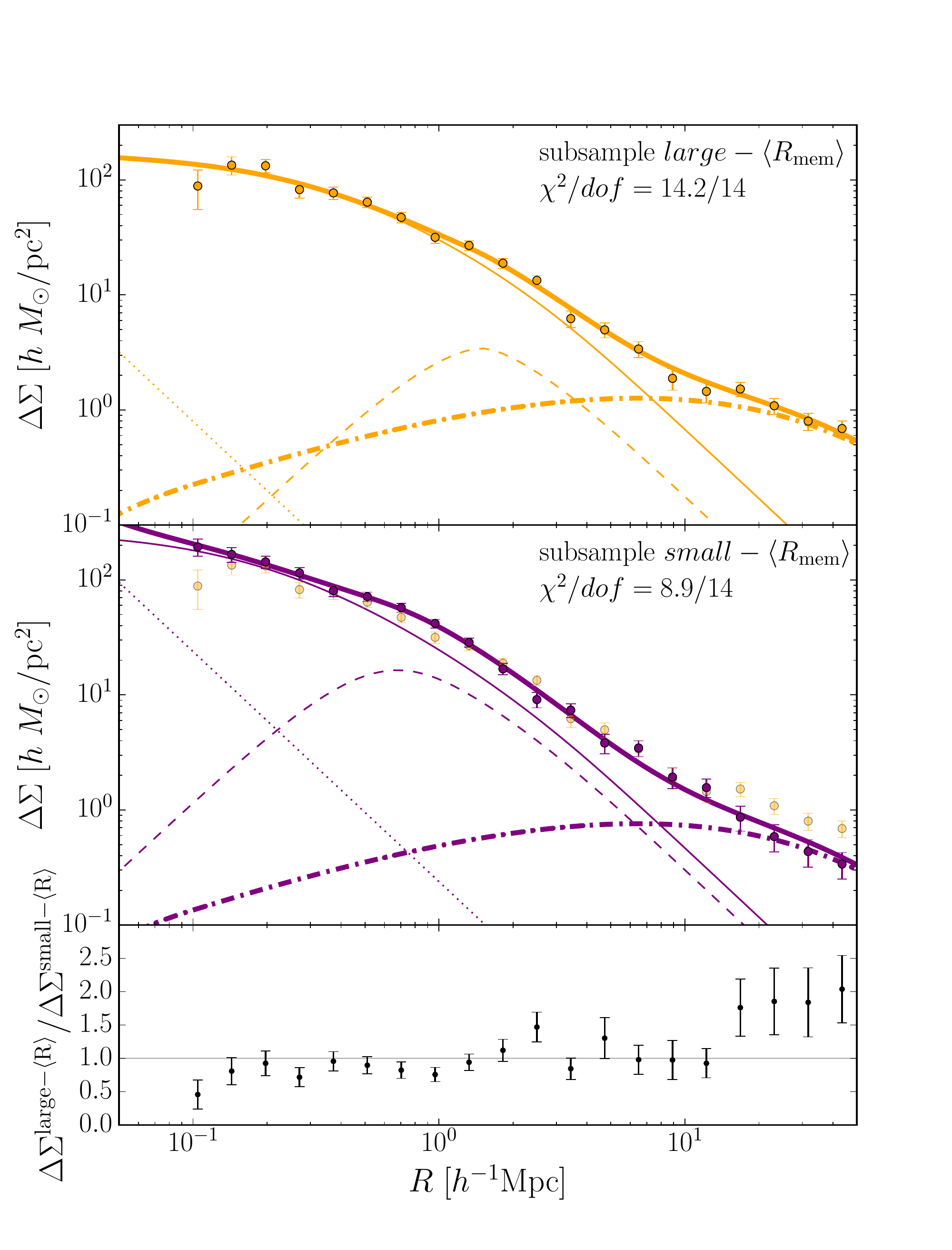}
 \caption{\label{fig:lensing_signal} Halo mass consistency and assembly
bias from the WL signal.: The data points with errorbars in the top
and middle panels show the excess surface mass density profile as a
function of the cluster-centric projected radius (in comoving units),
obtained from the WL measurements for the large- and small-$\ave{R_{\rm
mem}}$ subsamples of \redms clusters (see Eq.~\ref{eq:aveR}),
respectively.  The points from the top panel are reproduced in
semi-transparent color in the middle panel for comparison. The mass
profiles at small radii $R\simlt 10~h^{-1}{\rm Mpc}$ appear to have
similar amplitudes with slightly different shapes, but show a difference
in amplitude at $R\simgt 10~h^{-1}{\rm Mpc}$, as expected from {\it
assembly bias}.  The bold solid line shows the best-fit halo model, the
thin solid line is the centered 1-halo term, the dashed line is the
off-centered 1-halo term, while the dotted line corresponds to the
stellar mass contribution from the central galaxy.  Comparison between
the dot-dashed lines in the two panels implies that the 2-halo term
contributions, which arise from the average mass distribution
surrounding the clusters, are different by a factor of 1.6. The
bottom panel shows the ratios of the lensing signals, highlighting a
clear deviation from unity at $R\simgt 15~$Mpc$/h$.}
\end{figure}
\begin{figure}
\includegraphics[width=0.51\textwidth]{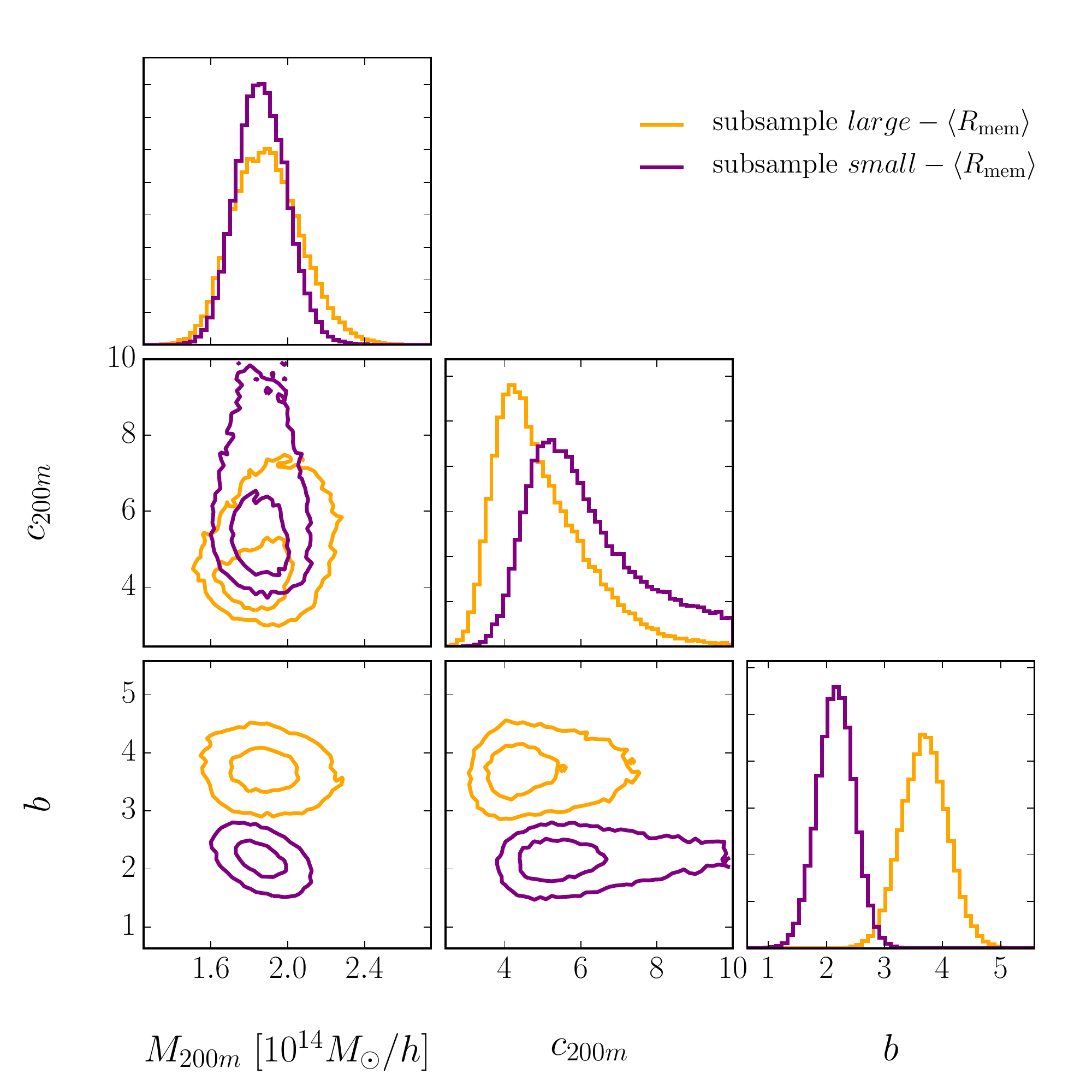}
 \caption{\label{fig:lensing_contour}
The posterior distributions of halo model parameters
given the WL signals for each of the two cluster subsamples shown in
Fig.~\ref{fig:lensing_signal}. The distributions include marginalization over
nuisance parameters which correspond to off-centering effects and stellar mass
contribution from the central galaxy (see text for details). The parameters
$M_{200m}$ and $c_{200m}$ are the halo mass and concentration parameters
that specify the average NFW profile of the clusters (1-halo term), while $b$ is
the linear halo bias of the cluster subsample. The posterior distributions show
the large- and small-$\ave{R_{\rm mem}}$ subsamples have similar halo mass, but
display a significant difference in their bias parameters.
 }
\end{figure}
{\it Weak lensing measurements --} The weak gravitational lensing (WL)
effect on the shapes of background galaxies can be used to measure the
average mass distribution around galaxy clusters. We use the shape catalog
of Ref.~\cite{Reyes:2012}, which is based on the photometric galaxy
catalog from the SDSS DR8 for this purpose. The galaxy shapes are measured by the
re-Gaussianization technique \cite{Hirata:2003}, and the systematic
uncertainties involved in the shape measurements have been investigated
in great detail in Ref.~\cite{Mandelbaum:2005}. The redshifts of source
galaxies are estimated based on the photo-$z$ code ZEBRA
\cite{Feldmann:2006, Nakajima:2012}.  The accuracy of the photometric redshift
is not crucial for our study, because the populations of source galaxies used to
compare the WL signals of our two cluster subsamples are identical. To measure
the cluster WL profiles, we use the same method as described in
Refs.~\cite{Mandelbaum:2013, Miyatake:2015}.

The top and middle
 panels of Fig.~\ref{fig:lensing_signal} show $\Delta
\Sigma(R)$, the excess surface mass density at a given projected radius
$R$ \cite{Mandelbaum:2005} for the large- and small-$\ave{R_{\rm mem}}$
subsamples of clusters, respectively.  The covariance matrix for each of
the measurements was estimated based on 83 jackknife regions of
approximately equal area covering the SDSS footprints (63 and 20 for the
northern and southern hemisphere footprints, respectively). The figure
shows that the WL signals of the two subsamples have very
similar amplitudes at small radii, $R\simlt 10~h^{-1}{\rm Mpc}$, and
consequently similar average halo masses.  However, the WL signals on
larger scales, $10\simlt R/[h^{-1}{\rm Mpc}]\le 50$, display a
significant difference, a signature of assembly bias, as explicitly
 shown in the bottom panel.

We perform halo model fits to the measurements of the WL signal of each
cluster subsample. Following Ref.~\cite{Hikage:2013}, we employ a simple
six parameter model fit to the WL signal,
\begin{eqnarray}
&&\Delta\Sigma(R;M_{200m},c_{200m},q_{\rm cen},
 \alpha_{\rm off}, M_\ast,b)\nonumber\\
&&\hspace{2em} = q_{\rm cen}\Delta\Sigma^{\rm NFW}(R;M_{200m},c_{200m})  \nonumber \\
&&\hspace{4em}  +(1-q_{\rm cen})\Delta\Sigma^{\rm NFW, off}(R;M_{200m},c_{200m},\alpha_{\rm
 off}) \nonumber \\
&&\hspace{4em}  +\Delta\Sigma^\ast(R;M_\ast)+\Delta \Sigma^{\rm 2-halo}(R;b).
\label{eq:halomodel}
\end{eqnarray}
The first term corresponds to the halo mass profile for the fraction
$q_{\rm cen}$ of clusters whose centers have been correctly identified,
while the second term corresponds to the clusters with misidentified
centers.  We assume that the halo mass profile is a smoothly truncated
version of the NFW profile \cite{Navarro:1996ApJ}, proposed in
Ref.~\cite{OguriHamana:2011}, specified by the halo mass and
concentration parameter, $M_{200m}$ and $c_{200m}$
\footnote{Throughout this paper we employ the halo mass and
  concentration definitions defined for the 200 overdensity with
  respect to the mean mass density}. We adopt $\tau_{200m}=2.6$ for
the smoothing kernel, although we confirmed that our result of similar
masses for the two subsamples is not sensitive to the chosen value of
$\tau_{200m}$.  We simply consider a single mass bin for host
halos. We assume that the normalized profile of the positions of
off-centered clusters with respect to their true center is given by
$u_{\rm off}(r)\propto \exp[-r^2/(2\alpha_{\rm off}^2R_{200m}^2)]$,
where $\alpha_{\rm off}$ describes the ratio of the off-centering
radius to $R_{200m}$. We also truncate the off-centering profile to
zero at $r>R_{200m}$.  The third term, $\Delta \Sigma_\ast\equiv
M_\ast/(\pi R^2)$, models a possible stellar mass contribution from
the central galaxies assuming a point mass. The fourth term
$\Delta\Sigma^{\rm 2-halo}(R)$ models the lensing contribution arising
from the two-point correlation function between the clusters and the
surrounding mass distribution. We employ the model given as $\Delta
\Sigma^{\rm 2h}(R)= b\int(k{\rm d}k/2\pi)\bar{\rho}_{\rm
  m}P^L_m(k;z_{\rm cl})J_2(kR)$, where $\bar{\rho}_{\rm m}$ is the
mean mass density today, $b$ is the linear bias parameter, and
$P^L_m(k;z_{\rm cl})$ is the linear mass power spectrum at the
averaged cluster redshift $z_{\rm cl}=0.24$, for the $\Lambda$CDM
model.

We explore the posterior distribution of the parameters given our
measurements using the Markov Chain Monte Carlo (MCMC) technique
\citep{Foreman-Mackey:2013}.  We use flat priors for all the parameters:
$M_{200m}/[10^{14}~h^{-1}M_\odot] \in [0.5, 50]$, $c_{200m} \in [1,
10]$, $q \in [0, 1]$, $\alpha_{\rm off} \in [10^{-4}, 1]$,
$M_\ast/[10^{12}~h^{-1}M_\odot] \in [0, 10]$, and $b \in [0, 10]$. In
Fig.~\ref{fig:lensing_contour}, we show the posterior distributions of
the parameters $\mtom$, $\ctom$ as well as $b$, comparing results for
the small- and large-$\ave{R_{\rm mem}}$ subsamples, after
marginalization over the off-centering parameters and the stellar mass
contribution \footnote{The off-centering parameters and the stellar mass
are very weakly constrained, e.g., at 68 percent confidence, $q_{\rm
cen}>0.7$ and $M_*<2.6\times10^{12}\msunh$ for both subsamples.}. The
halo masses are consistent with each other within the errorbars:
$M_{200m} / [10^{14}~h^{-1}M_\odot] = 1.87^{+0.12}_{-0.14}$ or
$1.88^{+0.16}_{-0.18}$ for the small- and large-$\ave{R_{\rm mem}}$
subsamples, respectively. The concentration parameters have strong
degeneracies with the off-centering parameters, but turn out to be
similar for the two subsamples after the marginalization. The halo bias
parameters are $b=2.17\pm 0.31$ and $3.67^{+0.40}_{-0.37}$,
respectively. The ratio $b^{{\rm large-}\ave{R_{\rm mem}}}/b^{{\rm
small-}\ave{R_{\rm mem}}}=1.64^{+0.31}_{-0.26}$, a 2.5$\sigma$ deviation from
unity.  For comparison, even if we take the halo masses for the two
subsamples at the extreme ends of their posterior distributions within
their 95\% C.L. interval, the halo bias model of Ref.~\cite{Tinker:2010}
predicts that the ratio is at most 1.13.

\begin{figure}
\includegraphics[width=0.5\textwidth]{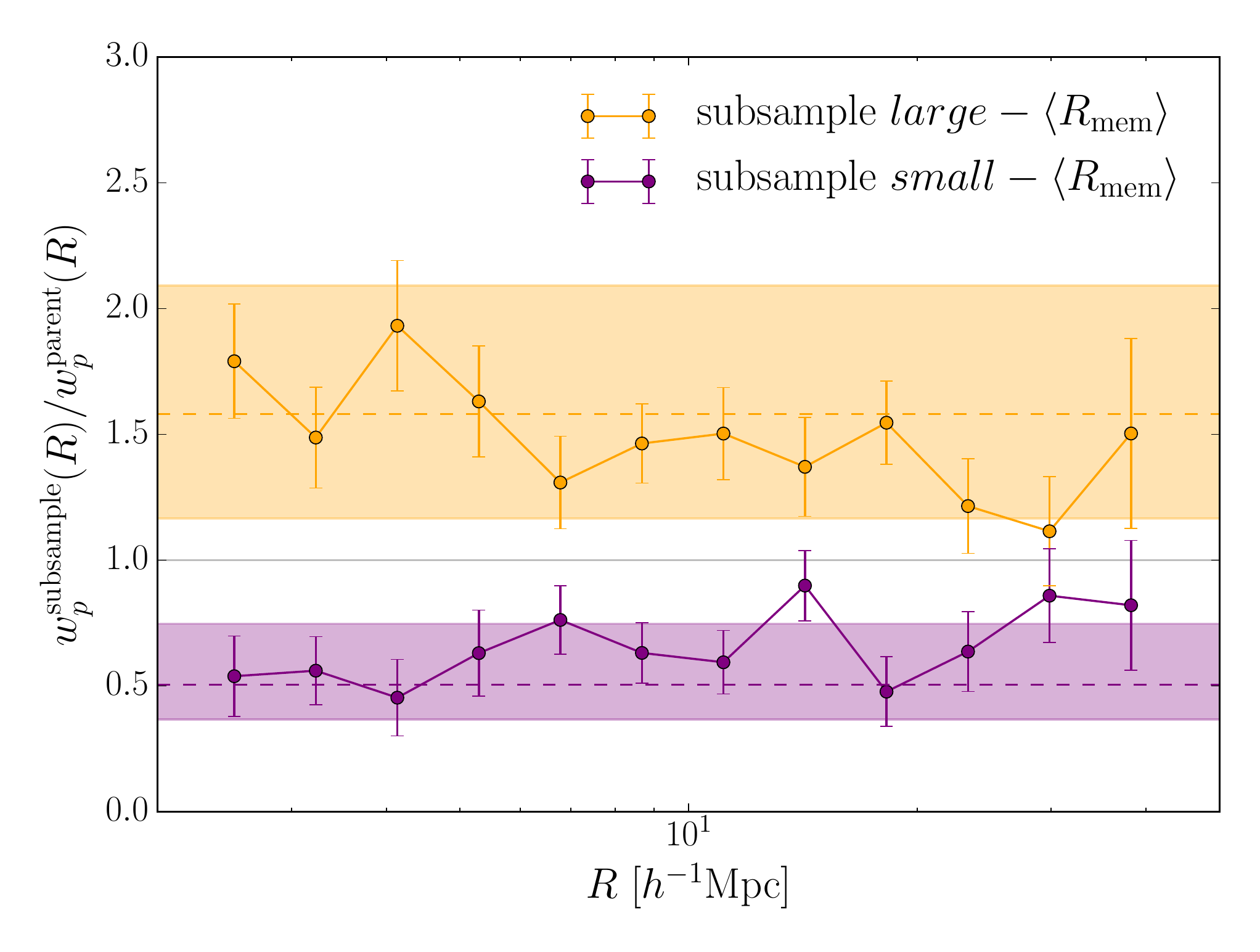}
 \caption{
Halo assembly bias from the projected clustering signal: The projected
auto-correlation function of clusters for each of the large- and
small-$\ave{R_{\rm mem}}$ subsamples, relative to that of the full sample (i.e.
 all clusters). The clustering signals at large separations,
$R\simgt 2~h^{-1}{\rm Mpc}$, show a significant difference, which is consistent
with the WL measurements in Fig.~\ref{fig:lensing_signal} as shown by the shaded
regions. 
 } 
 \label{fig:clustering_ratio}
\end{figure}
{\it Clustering measurements --} We now consider the auto-correlation functions
of clusters in the two subsamples to further confirm the difference in the
large-scale bias in the WL signals. To avoid
redshift-space distortions, we consider the projected correlation function
\begin{equation}
w_p(R)=2\int_0^{\Pi_{\rm max}=100~h^{-1}{\rm Mpc}} \rmd\Pi~ \xi(R,\Pi)\,,
\end{equation}
where $R$ and $\Pi$ are the projected and line-of-sight separations between
cluster pairs, and $\xi(R,\Pi)$ is the three-dimensional correlation function.
We compute the three-dimensional correlation function $\xi(R,\Pi)$ using the
Landy \& Szalay estimator \cite{Landy:1993}, and replace the integral over
line-of-sight by discrete summation with $\Delta \Pi=1~ h^{-1}{\rm Mpc}$.

In Fig.~\ref{fig:clustering_ratio}, we show the ratio of the projected
auto-correlation functions measured from our subsamples, relative to
that of the parent sample (i.e., all the clusters), along with jackknife
error estimates.  Over the range of separations we have considered, the
two subsamples show significantly different clustering amplitudes than
the parent sample, giving an independent confirmation of assembly bias.
To quantify the significance, we fit a constant parameter model,
$\zeta_0$, to $\zeta(R)=\sqrt{w_{p}^{\rm large-\ave{R_{\rm mem}}}/w_{p}^{\rm
small-\ave{R_{\rm mem}}}}$, accounting for the covariance between the data
points.  We obtain $\zeta_0=1.40\pm0.09$ ($1.34\pm0.12$ if we
restrict to $R>10\mpch$), corresponding to a $4.4\sigma$ deviation
from $\zeta_0=1$. Using both the ratios of the clustering signal and the lensing
signals at the three outermost bins in Fig.~\ref{fig:lensing_signal},
we obtain $\zeta_0=1.41\pm0.09$ with the significance of
4.7$\sigma$, after accounting for the cross-covariance between the
clustering and lensing signals. To be explained by halo mass
alone, the bias ratio of 1.4 (1.6) implied from the clustering (lensing)
measurements requires a factor of 3 (4) difference in the halo masses,
which is disfavored by the 1-halo lensing signals.

The shaded regions in Fig.~\ref{fig:clustering_ratio} show that the
difference from the clustering amplitudes is consistent with the WL
measurements, within the errorbars. The shaded regions were obtained
with the results of fits to the lensing signals of the subsamples and
the full sample as in Fig.~\ref{fig:lensing_contour}. They show the
posterior distribution of the square of bias ratio of each subsample to
the full sample, after marginalization over other parameters.  The
dashed lines correspond to the value of the maximum posterior
probability.

{\it Conclusion and discussion} -- A combination of WL and clustering
measurements of \redms galaxy clusters show evidence of the assembly
bias for cluster-sized halos with mass $\sim2\times10^{14}\msunh$, with
the significance level of 2.5$\sigma$ and 4.4$\sigma$ for each of two
observables. One of the implications of this measurement is that the
halo mass based on large-scale clustering amplitude is sensitive to the
cluster selection function, which should be taken into account for
precision cosmology.

Could the large difference in halo bias arise from orientation biases in
the cluster identification algorithm? Could the small-$\ave{R_{\rm
mem}}$ clusters preferentially reside in smaller mass halos and
consequently have smaller bias, but their weak lensing mass is large due
to a filament aligned along the line-of-sight?  For the halo bias
difference to arise just from a halo mass difference, the subsamples
have to differ in halo mass by a factor of $\sim3-4$, which is quite
unlikely to occur just by projection effects \cite[see
e.g.,][]{Dietrich:2014, Becker:2011}. 
Clusters affected by
projection (intrinsically lower mass systems) are expected to cause an
increase, contrary to a decrease in $\ave{R_{\rm mem}}$
\cite{Cohn:2007}. 
We have explicitly verified that the line-of-sight velocity dispersions of
member galaxies around the two cluster subsamples are very similar as well.
Furthermore, our preliminary investigations indicate that a number of properties
of central galaxies (such as their stellar masses, stacked spectra, velocity
dispersions) in the two subsamples also do not show any significant differences.

It is worth exploring whether the amplitude of the observed assembly
bias is consistent with predictions in $\Lambda$CDM cosmologies.
Ref.~\cite{Wechsler:2006} used $N$-body simulations to study the halo
assembly bias in $\Lambda$CDM models, and found that the bias difference
for rare objects such as our clusters, when subdivided by the halo
concentration, is $\sim1.25$, somewhat smaller than our finding
(difference of $1.40 \pm 0.09$ for the clustering measurement). However,
there are differences in our method: we used the distribution of member
galaxies (which should correspond to subhalo locations) for the
subdivision, the lensing measurements do not show a clear difference in
the halo concentration for the two subsamples
(Fig.~\ref{fig:lensing_contour}), and the scales we consider include the
mildly non-linear regime. Our preliminary analysis using a high-resolution
$N$-body simulation, indicates qualitative agreement between the sign and
strength of the signal in the simulations and in the data. These results will be
presented elsewhere \cite{More:2015}.

A proper comparison between data and simulations will require a careful
study of the cluster selection function, as well as observational
effects due to line-of-sight projections and the photo-$z$
uncertainties, which can be studied by running the same cluster-finding
algorithm on realistic simulated galaxy catalogs with full halo assembly
histories. Additionally, on the
observational side, one could improve the signal-to-noise ratio of the
measurement by using the cross-correlations of the cluster samples of this study
with a sample of galaxies such as a catalog of luminous red galaxies that has a
much higher spatial number density. We defer these topics to a study in the near
future.

\begin{acknowledgments}
We thank Tom Abel, Neal Dalal, Oliver Hahn, Benedikt Diemer, Andrey
Kravtsov for enlightening discussions. We also thank the three
anonymous referees for a careful reading of the manuscript and their
suggestions. HM is supported in part by Japan Society for the
Promotion of Science (JSPS) Research Fellowships for Young Scientists
and by the Jet Propulsion Laboratory, California Institute of
Technology, under a contract with the National Aeronautics and Space
Administration. MT and SM are supported by World Premier
International Research Center Initiative (WPI Initiative), MEXT,
Japan, and by the FIRST program `Subaru Measurements of Images and
Redshifts (SuMIRe)', CSTP, Japan. SM and MT are also supported by
Grant-in-Aid for Scientific Research from the JSPS Promotion of
Science (No. 15K17600, 23340061, and 26610058), MEXT Grant-in-Aid for
Scientific Research on Innovative Areas (No.  15H05893) and by JSPS
Program for Advancing Strategic International Networks to Accelerate
the Circulation of Talented Researchers.  DNS is partially supported
by the NSF AST-1311756 and NASA NNX14AH67G.  RM acknowledges the
support of the Department of Energy Early Career Award program. ESR is
partially supported by the U.S. Department of Energy contract to SLAC
no. DE-AC02-76SF00515.

Funding for SDSS-III has been provided by the Alfred P. Sloan
Foundation, the Participating Institutions, the National Science
Foundation, and the U.S. Department of Energy Office of Science. The
SDSS-III web site is http://www.sdss3.org/.

SDSS-III is managed by the Astrophysical Research Consortium for the
Participating Institutions of the SDSS-III Collaboration including the
University of Arizona, the Brazilian Participation Group, Brookhaven
National Laboratory, Carnegie Mellon University, University of Florida,
the French Participation Group, the German Participation Group, Harvard
University, the Instituto de Astrofisica de Canarias, the Michigan
State/Notre Dame/JINA Participation Group, Johns Hopkins University,
Lawrence Berkeley National Laboratory, Max Planck Institute for
Astrophysics, Max Planck Institute for Extraterrestrial Physics, New
Mexico State University, New York University, Ohio State University,
Pennsylvania State University, University of Portsmouth, Princeton
University, the Spanish Participation Group, University of Tokyo,
University of Utah, Vanderbilt University, University of Virginia,
University of Washington, and Yale University.

\end{acknowledgments}

 \bibliography{paper}

\end{document}

%% file: commands.tex
\newcommand{\mpch}{\>h^{-1}{\rm {Mpc}}}

\newcommand{\msunh}{\>h^{-1}\rm M_\odot}

\def\gcm3{\mathrm{g} / \mathrm{cm}^3}

\def\gtsima{$\; \buildrel > \over \sim \;$}
\def\ltsima{$\; \buildrel < \over \sim \;$}
\def\prosima{$\; \buildrel \propto \over \sim \;$}
\def\gsim{\lower.7ex\hbox{\gtsima}}
\def\lsim{\lower.7ex\hbox{\ltsima}}
\def\simgt{\lower.7ex\hbox{\gtsima}}
\def\simlt{\lower.7ex\hbox{\ltsima}}
\def\simpr{\lower.7ex\hbox{\prosima}}

\def\rmd{{\rm d}}